\documentclass[twocolumn,amsmath,amssymb,aps,pre,showpacs]{revtex4}
\usepackage{graphics}

\begin{document}

\title{Fluctuation formula in the Nos\'e-Hoover thermostated Lorentz gas}

\author{M. Dolowschi{\'a}k} \email{dolowsch@szerenke.elte.hu} 
\author{Z. Kov{\'a}cs} \email{kz@garfield.elte.hu}

\affiliation{Institute for Theoretical Physics, E{\"o}tv{\"o}s University,
          Pf.\ 32, H--1518 Budapest, Hungary}

\begin{abstract}
In this paper we examine numerically the Gallavotti-Cohen fluctuation 
formula for phase-space contraction rate and entropy production rate 
fluctuations in the Nos\'e-Hoover thermostated periodic Lorentz gas. 
Our results indicate that while the phase-space contraction rate 
fluctuations violate the fluctuation formula near equilibrium states, 
the entropy production rate fluctuations obey this formula near and far from 
equilibrium states as well. 
\end{abstract}


\maketitle

\section{Introduction}


In recent years a large number of papers focused on various 
fluctuation formulas (FFs) with both theoretical and numerical tools, and 
today these seem to be one of the most interesting results in the 
field of statistical physics of nonequilibrium systems \cite{ftreview}.
This behavior was observed numerically in a system of 
thermostated fluid particles undergoing shear flow \cite{shearflow}.
It was an important property of the FF that it seemed to be valid for 
large external forcings as well \cite{chaodyn}, therefore it was 
considered that it could shed some light on the thermodynamical behavior 
of systems far from equilibrium. Consequently, significant theoretical 
efforts have been made to find a common property behind the observed FF 
and these efforts resulted in various fluctuation theorems (FTs): the 
Gallavotti-Cohen approach built on the chaotic hypothesis 
\cite{gcft,dynens}, the Evans-Searles theorems \cite{es},
deterministic local FT \cite{revdamp},
the FT for stochastic systems \cite{kurchan,es4,lebspo},
the theorem of Maes established on the Gibbs Property \cite{maes}, and
the FT for open systems \cite{tel}. The Gallavotti-Cohen FT/FF serves as the 
basis of the numerical investigations presented in this paper.

All of the applied theoretical methods shared the property of 
putting extra presumptions on the physical systems (i.e., chaoticity, 
stochasticity) that could not be proved {\em a priori}. 
This situation naturally raised the need to study the FF numerically 
and compare the numerical results to the theoretical predictions.
Up to now several physical models have been investigated numerically, such as
the two-dimensional (2D) reversibly damped fluids \cite{revdamp}, 
the chains of weakly
interacting cat maps \cite{catmap}, the Fermi-Ulam-Pasta chain 
\cite{anhlatt}, and the periodic Lorentz Gas (PLG) thermostated by the 
Gaussian isokinetic (GIK) thermostat \cite{exptest,flform}. 

The FF is a symmetry property of the probability density function (PDF) of
a dynamically measured quantity $\pi$ connecting the probabilities of 
measuring $\pi$ values with equal magnitudes but opposite signs. More 
precisely, let $\pi_\tau(t)$ denote the quantity $p$ 
averaged over a time interval of length $\tau$ centered around time $t$:
$\pi_\tau(t)= \frac{1}{\tau}
      \int\limits_{-{\tau}/{2}}^{{\tau}/{2}}p(t+t')\,\mathrm{d}t'$. 
Considering it as a stochastic variable $x$, its statistical properties in
a steady state can be characterized by the PDF $\Pi_\tau(x)$. The FF 
states that the PDF $\Pi_\tau(x)$ has the following property:
\begin{equation}
  \label{flform}
  \lim_{\tau\to\infty}
        \frac{1}{\tau} \ln \frac{\Pi_\tau(x)}{\Pi_\tau(-x)}=x\,. 
\end{equation}
In other words, for large enough $\tau$ values the probability of 
observing $-\pi_\tau$ is exponentially smaller than the probability of 
observing $\pi_\tau$ .

In \cite{exptest,flform} the examined physical quantity $\pi_\tau(t)$ was 
the phase-space contraction rate (PSCR), which due to this special 
property of the applied GIK thermostat was equal to the thermodynamical 
entropy production rate (EPR) at any given time. 
However, this identity does not hold in general \cite{epr}, and 
PSCR and EPR fluctuations can have different PDFs as, e.g., in the case of 
Nos\'e-Hoover thermostated systems (see Sec.~\ref{secsys}).
The theoretical methods using the Sinai-Ruelle-Bowen (SRB) measure 
\cite{gcft,shearflow} to calculate the probabilities of trajectory 
segments predict that the quantity obeying the FF is the PSCR and further 
nontrivial theoretical efforts are needed to establish similar results for
the EPR fluctuations \cite{epr}.

The main purpose of examining the Nos\'e-Hoover (NH) thermostated PLG 
is to investigate which of the above-mentioned physical 
quantities obey the FF in a system, where the corresponding PDFs are not 
identical.
In addition to this, checking the NH thermostated PLG numerically against 
the FF is in itself an important task, given that the NH thermostat is 
one of the two generally used dynamical thermostats; 
the transport properties of this model have been recently investigated in 
\cite{nosehoover}.
In Sec.~\ref{secsys} we describe the examined model;
in Sec.~\ref{secnumres} we present our numerical results; and in 
Sec.~\ref{seccon} we summarize our conclusions.


\section{The System}
\label{secsys}

One of the most investigated models suitable for studying 
transport phenomena is the field-driven thermostated periodic Lorentz Gas.
This model consists of a charged particle subjected to external 
electric field and moving in the lattice of elastic scatterers.
Due to the applied electric field, one must use a thermostating mechanism 
to achieve a steady state in the system; such a tool is a dynamical 
thermostat. 
Two types of dynamical thermostats have been applied to the PLG up to now:
the GIK thermostat producing microcanonical distribution \cite{gik} and 
the Nos\'e-Hoover thermostat producing canonical distribution in 
equilibrium \cite{nosehoover,thermostats}.

We present the equations of motion of the two-dimensional NH
thermostated PLG in dimensionless form:
mass and electric charge are measured in units of the particle's mass $m$
and charge $q$ and the unit length is chosen to be the radius of the 
scatterers ($R=1$). 
Let ${\bf q}=\left(q_1,q_2\right)$ denote the position and
${\bf p}=\left(p_1,p_2\right)$ the momentum of the particle, 
then the phase-space vector of the system is 
${\mathbf \Gamma}=\left({\bf q},{\bf p},\zeta \right)$, 
where $\zeta$ is the state variable of the thermal reservoir.
Between two subsequent collisions the state of the system is evolved 
smoothly by the differential equation
\begin{eqnarray}
  \label{diffeq}
   \dot{\bf q} & = & {\bf p}\,,       \nonumber \\
   \dot{\bf p} & = & {\bf E} - \zeta {\bf p}\,, \\
   \dot \zeta   & = & \frac{1}{\tau^2_{resp}} \left( \frac{p^2}{2T} - 1 \right)\,, \
\nonumber
\end{eqnarray}
and is transformed abruptly at every elastic collision. In this equation 
${\bf E}$ is the the external electric field, $\tau_{resp}$ is the 
response time of the reservoir, and $T$ is the temperature satisfying 
$\left<p^2\right>=2T$.

In the simulations presented in this paper we have used a
{\em square lattice} of circular scatterers, however, we have investigated 
numerically other lattices as well (e.g, triangular), but have not found any
relevant differences concerning the results presented in 
Sec.~\ref{secnumres}.

Energy dissipation can be measured by the phase-space contraction rate 
$\sigma$ and can be computed by taking the divergence of the right-hand 
side of Eq.~(\ref{diffeq}) as:
\begin{eqnarray}
  \label{sigma}
   \sigma(t)= - \mbox{div}\, {\bf \dot\Gamma}(t) =\zeta(t)\,.
\end{eqnarray}
The entropy production rate $\xi$ can be formally defined by the 
expression of irreversible thermodynamics 
\begin{eqnarray}
  \label{ksi}
  \xi(t)=\frac{{\bf J}(t){\bf E}}{T}=\frac{{\bf p}(t){\bf E}}{T}\,, 
\end{eqnarray}
where $T$ is the kinetic temperature. We note that this quantity is 
identical to the dissipation function of the Evans-Searles theorem 
\cite{ftreview}. It can be shown that in this model
$\left<\sigma\right>=\left<\xi\right>$, however, the identity 
$\sigma(t)=\xi(t)$ does not hold at all times, as opposed to the case of 
the GIK thermostated PLG.


\section{Numerical Results}
\label{secnumres}

The objective of the numerical simulation is to measure the PDFs 
$\Sigma_\tau(x)$ and $\Xi_\tau(x)$ of the averaged quantities 
$\sigma_\tau(t)=\frac{1}{\tau}\int_{-\frac{\tau}{2}}^{\frac{\tau}{2}} \
\zeta(t^{'})\,\mathbf{d} t^{'}$ and $\xi_\tau(t)=\frac{1}{\tau}\
\int_{-\frac{\tau}{2}}^{\frac{\tau}{2}}\
\frac{{\bf p}(t+ t^{'}){\bf E}}{T}\,\mathbf{d} t^{'}$
and check the validity of the FF for them. In order to perform 
this task we should evolve the state of the system $\bf{\Gamma}$ along a 
long trajectory, which requires the algorithm to be very efficient.
This need motivated us to implement an {\em event driven algorithm} 
that generates and handles events, such as the collision of the
particle with a scatterer and the replacement of the particle from one 
simulation cell into the other. 
The most sensitive issue when applying such an algorithm is to 
determine the point of time when a specific event occurs; in computer 
science this problem is known as {\em collision detection}.
Since in the case of the Nos\'e-Hoover thermostat the velocity of the 
particle is not upper bounded, we could not have chosen the simplest such
method, the so-called naive algorithm, which could have been applied in 
the case of the GIK thermostat. 
Instead of this we have applied the method of building the time estimation
into the Runge-Kutta integrator, which is used in simulating particle 
laden flows and coupled particle-field systems (see Ref.~\cite{colldet}).
\begin{figure}[h]
\scalebox{0.3}{\rotatebox{-90}{\includegraphics{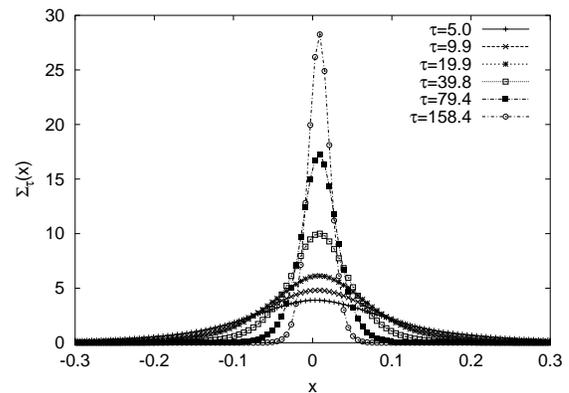}}}
\caption{The probability density function $\Sigma_\tau(x)$ of the \
averaged phase space contraction rate fluctuations $\sigma_\tau$ in a \
configuration \
close to equilibrium $\left( {\bf E}=\left(0.1,0.2\right) \right)$. }
\label{fighistsigmaeq}
\end{figure}
\begin{figure}[!h]
\scalebox{0.3}{\rotatebox{-90}{\includegraphics{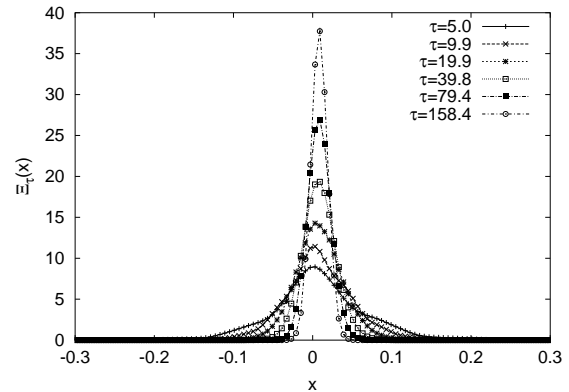}}}
\caption{The probability density function $\Xi_\tau(x)$ of the averaged \
entropy production rate fluctuations $\xi_\tau$ in a configuration close \
to equilibrium $\left({\bf E}=\left(0.1,0.2\right)\right)$.}
\label{fighistksieq}
\end{figure}

The PDFs $\Sigma_\tau(x)$ and $\Xi_\tau(x)$ were constructed by periodically 
computing the quantity $\sigma_\tau(t)$ and $\xi_\tau(t)$  
along a long particle trajectory and making a histogram of the computed 
data. 
In building the histogram we have used overlapping and nonoverlapping 
windows techniques as well, but have not found any relevant differences
between them concerning the results of this paper. Throughout the
presented numerical experiments we have used the $T=1.0$ and 
$\tau_{resp}=1.0$ 
values, however we have tested several other configurations as well. 
We have simulated $t=10^7$ long particle trajectories resulting in 
approximately $2 \times 10^6$ collisions with the scatterers.

\begin{figure}[!h]
\scalebox{0.3}{\rotatebox{-90}{\includegraphics{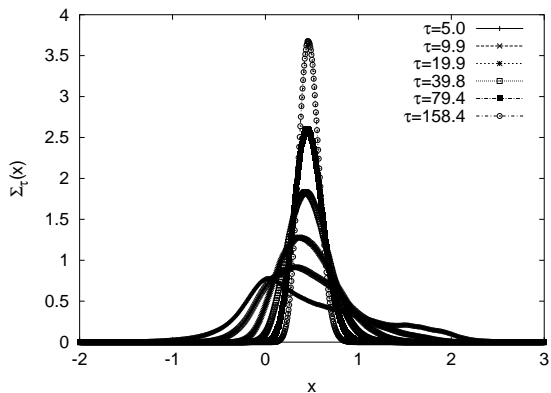}}}
\caption{The probability density function $\Sigma_\tau(x)$ of the \
averaged phase space contraction rate fluctuations $\sigma_\tau$ in a \
configuration far from equilibrium \
$\left({\bf E}=\left(0.8,1.6\right)\right)$]. }
\label{fighistsigmanoneq}
\end{figure}

\begin{figure}[!h]
\scalebox{0.3}{\rotatebox{-90}{\includegraphics{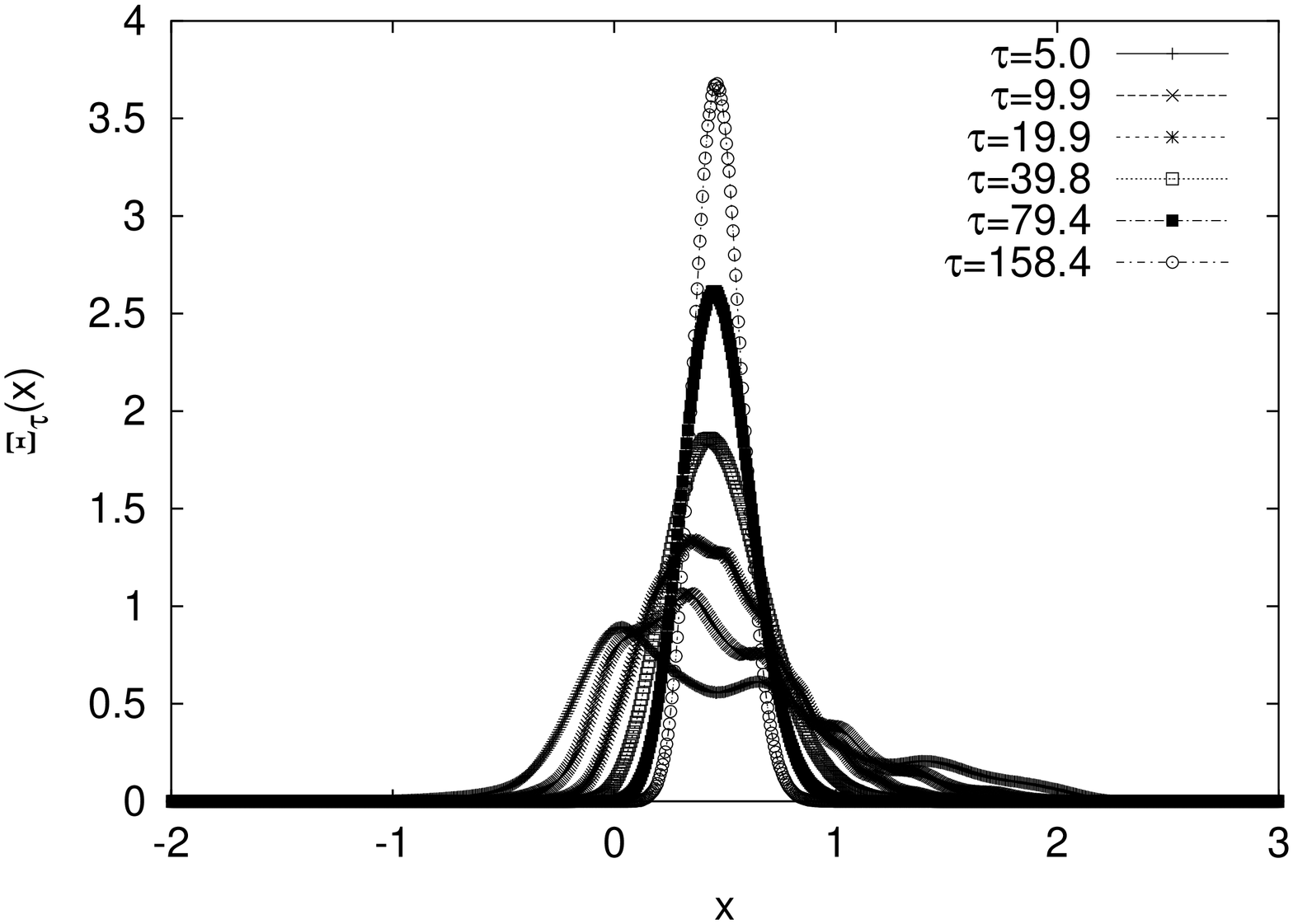}}}
\caption{The probability density function $\Xi_\tau(x)$ of the averaged \
entropy production rate fluctuations $\xi_\tau$ in a configuration far \
from equilibrium $\left({\bf E}=\left(0.8,1.6\right)\right)$.}
\label{fighistksinoneq}
\end{figure}

Figures \ref{fighistsigmaeq}, \ref{fighistksieq}, \ref{fighistsigmanoneq} 
and \ref{fighistksinoneq} show the functional form of the PDFs 
$\Sigma_\tau(x)$ and $\Xi_\tau(x)$ for small and large external fields. 
Examining the figures one can make the following interesting observations:
\begin{enumerate}

\item For low $\left|{\bf E}\right|$ values (close to equilibrium) 
in Figs \ref{fighistsigmaeq} \ref{fighistksieq} the PDF 
$\Sigma_\tau(x)$ seems to be more symmetric than the PDF $\Xi_\tau(x)$.

\item For low $|{\bf E}|$ values in Fig.~\ref{fighistsigmaeq} and 
Fig.~\ref{fighistksieq} the PDF $\Xi_\tau(x)$ seems strictly narrower 
than $\Sigma_\tau(x)$; indeed in the $|{\bf E}| \to 0$ limit 
$\Xi_\tau(x)$ should converge to a Dirac $\delta$ function .
[$\xi_\tau(t)=0$],
opposed to $\sigma_\tau$ that can fluctuate even in equilibrium.

\item For high $\tau$ values in Fig.~\ref{fighistsigmaeq}, 
\ref{fighistksieq}, \ref{fighistsigmanoneq} and \ref{fighistksinoneq} the 
curves seem to be indistinguishable from a Gaussian; 
indeed fitting a Gaussian onto the measured values yields an excellent
visual agreement.
\end{enumerate}
\begin{figure}[!th]
\scalebox{0.3}{\rotatebox{-90}{\includegraphics{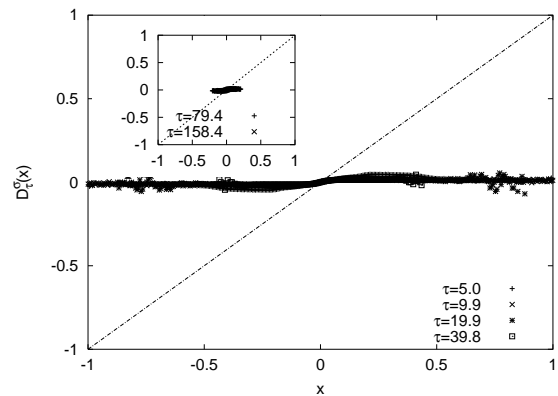}}}
\caption{The quantity $D^{\sigma}_\tau(x)$ for the phase space contraction 
rate in a configuration close to equilibrium \
$\left({\bf E}=\left(0.1,0.2\right)\right)$.}
\label{figftsigmaeq}
\end{figure}
\begin{figure}[!th]
\scalebox{0.3}{\rotatebox{-90}{\includegraphics{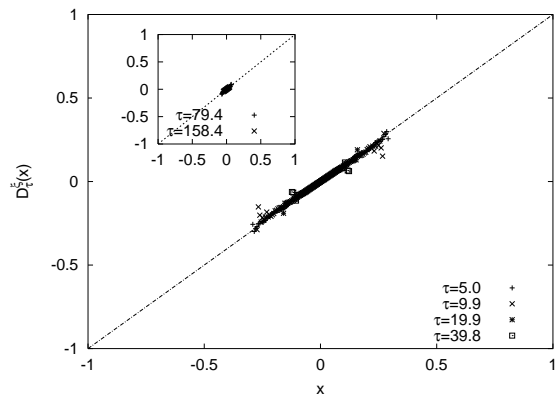}}}
\caption{The quantity $D^{\xi}_\tau(x)$ for the entropy production rate 
in a configuration close to equilibrium \
$\left({\bf E}=\left(0.1,0.2\right)\right)$. }
\label{figftksieq}
\end{figure}
\begin{figure}[!th]
\scalebox{0.3}{\rotatebox{-90}{\includegraphics{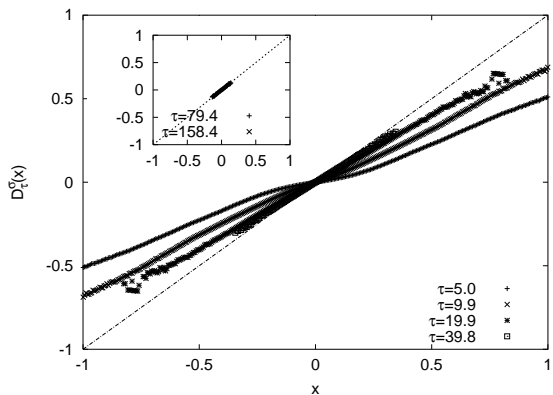}}}
\caption{ The quantity $D^{\sigma}_\tau(x)$ for the phase space contraction 
rate in a configuration far from equilibrium \
$\left({\bf E}=\left(0.8,1.6\right)\right)$.}
\label{figftsigmanoneq}
\end{figure}
\begin{figure}[!h]
\scalebox{0.3}{\rotatebox{-90}{\includegraphics{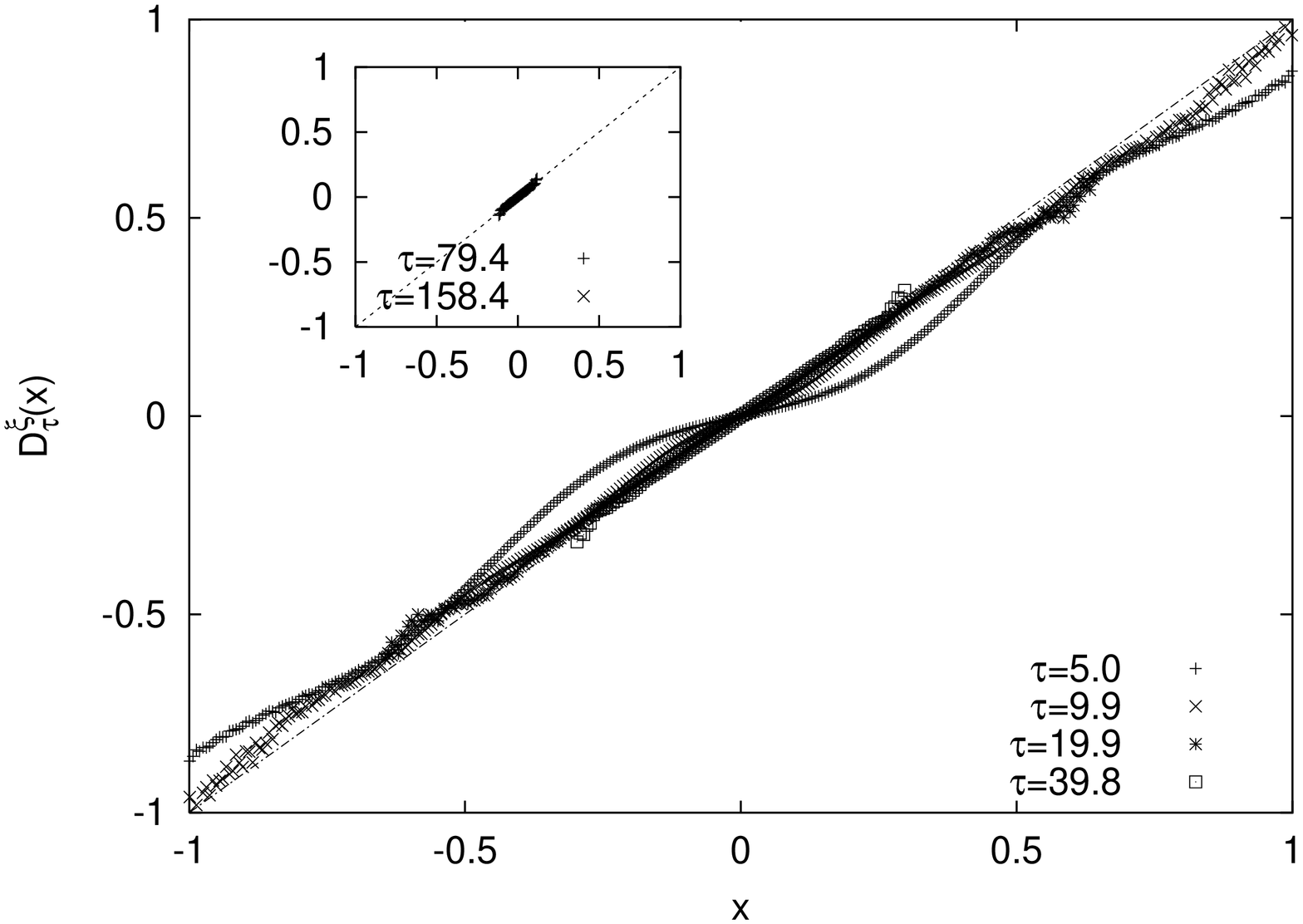}}}
\caption{The quantity $D^{\xi}_\tau(x)$  for the entropy production rate \
in a configuration far from equilibrium \
$\left({\bf E}=\left(0.8,1.6\right)\right)$.}
\label{figftksinoneq}
\end{figure}

With the measured values of $\Sigma_\tau(x)$ and $\Xi_\tau(x)$ one 
can check the FF in this model. In order to visualize the FF we may 
introduce the quantities
\begin{eqnarray}
\label{Dtau}
D^{\sigma}_\tau(x)&=&\frac{1}{\tau}\,\ln \frac{\Sigma_\tau(x)}{\Sigma_\tau(-x)}\,, \\
D^{\xi}_\tau(x)&=&\frac{1}{\tau}\,\ln \frac{\Xi_\tau(x)}{\Xi_\tau(-x)}\,,
\end{eqnarray}  
which are shown on Fig.~\ref{figftsigmaeq}, \ref{figftksieq}, 
\ref{figftsigmanoneq} and \ref{figftksinoneq}.
With these quantities Eq.~(\ref{flform}) reads as 
$\lim_{\tau \to \infty}D^{\sigma}_\tau(x)=x$ and
$\lim_{\tau \to \infty}D^{\xi}_\tau(x)=x$.
Examining Fig.~\ref{figftsigmaeq}, \ref{figftksieq}, 
\ref{figftsigmanoneq} and \ref{figftksinoneq}, one can
conclude that
\begin{enumerate}
\item For small external forcings and for numerically available $\tau$ 
values the $\sigma_\tau$ fluctuations seem to violate the FF 
(Fig.~\ref{figftsigmaeq}); this observation is supported by the fact that
any symmetric function substituted in Eq.~(\ref{Dtau}) yields zero, and
due to the time-reversing symmetry of Eq.~(\ref{diffeq}) the PDF 
$\Sigma_\tau(x)$ is expected to be close to a symmetric function for small 
external forcings (see Ref.~\cite{thermsys}).
\item The $\xi_\tau$ fluctuations seem to obey the FF both for small and 
large external forcings. 
\item As the external forcing grows, the $\sigma_\tau$ fluctuations seem 
to obey the FF for large $\tau$ values (Fig.~\ref{figftsigmanoneq}) 
similarly to $\xi_\tau$ fluctuations. 
\end{enumerate}

We note that we have examined several other configurations and have 
found no significant qualitative differences in the observed behavior of 
the PDFs.


\section{Conclusion}
\label{seccon}

In this paper we have presented numerical evidence showing that the 
phase-space contraction rate fluctuations violate the fluctuation formula 
[Eq.~(\ref{flform})] in or close to equilibrium in the Nos\'e-Hoover 
thermostated periodic Lorentz gas. This observation is completely in line 
with the theoretical predictions of Evans et al in Ref.~\cite{thermsys}.

On the other hand, we also demonstrated that the entropy production rate 
fluctuations satisfy the fluctuation formula, which is by no means 
trivial. It should also be noted that from the physical point of view the 
entropy production rate is the more relevant quantity due to its relation 
to thermodynamics and the availability for measurement in physical 
experiments. A direct consequence of our results is that the phase-space 
contraction rate and entropy production rate cannot be treated as 
interchangeable qunatities in fluctuation formulas, therefore statements
regarding the applicability of the FF for various models should always 
clarify which fluctuations they refer to.

\begin{acknowledgments}
The authors are grateful to Rainer Klages and Tam{\'a}s T{\'e}l for 
fruitful discussions and careful reading of the manuscript. 
This work was supported by the Hungarian Academy of Sciences and by the
Hungarian Scientific Research Foundation (Grant No.\ OTKA T032981).
\end{acknowledgments}



\begin{thebibliography}{99}

\bibitem{ftreview} For a review on the topic see
D.J. Evans and D.J. Searles, Adv. Phys. {\bf 51}, 1529 (2002) 

\bibitem{shearflow} D.J. Evans, E.G.D. Cohen and G.P. Morriss, Phys. Rev. Lett. {\bf 71}, 2401 (1993)

\bibitem{chaodyn} For an overview, see G. Gallavotti, Chaos 8, 384 (1998) 
 and references therein.

\bibitem {gcft} G. Gallavotti and E.G.D. Cohen, J. Stat. Phys. {\bf 80}, 931 (1995) 
\bibitem{dynens} G. Gallavotti and E.G.D. Cohen, Phys. Rev. Lett. {\bf 74}, 2694 (1995)

\bibitem{es} D.J. Evans and D.J. Searles, Phys. Rev. E {\bf 50}, 1645 (1994),
D.J. Evans and D.J. Searles, Phys. Rev. E {\bf 52}, 5839 (1995),
D.J. Evans and D.J. Searles, Advances in Phys.,{\bf 51}, 1529 (2002).

\bibitem{revdamp} Giovanni Gallavotti, Lamberto Rondoni and Enrico Segre, Physica D {\bf 187}, 338 (2004).

\bibitem{kurchan} J. Kurchan, J. Phys. A, {\bf 31}, 3719 (1998) 

\bibitem{es4} D.J. Searles and D.J. Evans, Phys. Rev. E {\bf 60}, 159 (1999)

\bibitem{lebspo} J. L. Lebowitz and H. Spohn,J. Stat. Phys., {\em 95},333 (1999)

\bibitem{maes} C. Maes, J. Stat. Phys, 95, 367 (1999)

\bibitem{tel} L. Rondoni, T. T{\'e}l and J. Vollmer, Phys. Rev. E {\bf 61}, 4679 (2000)


\bibitem{anhlatt} S.Lepri, A. Politi and R. Livi, Physica D {\bf 119}, 140 (1998)


\bibitem{catmap} Giovanni Gallavotti, Fabio Perroni, e-print chao-dyn/9909007,

\bibitem{exptest}F. Bonetto, G. Gallavotti and P.L. Garrido, Physica D {\bf 105}, 226 (1997)

\bibitem{flform} M. Dolowschi{\'a}k and Z. Kov{\'a}cs, Phys. Rev. E {\bf 66}, 066217 (2002)


\bibitem{epr} R. Klages, e-print nlin.CD/0309069 (2003) and
E.G.D. Cohen and L. Rondoni, Chaos {\bf 8}, 357 (1998)


\bibitem{nosehoover} K. Rateitschak, R. Klages and W.G. Hoover, J. Stat. Phys. {\bf 101}, 61-77 (2000)

\bibitem{gik} C.P. Dettmann and G.P. Morriss, Phys. Rev. E {\bf 53}, 2495 (1996)  


\bibitem{thermostats} For a review of deterministic thermostats, 
 see R. Klages, e-print nlin.CD/0309069 (2003)
 and G.P. Morriss and C.P. Dettmann, Chaos {\bf 8}, 321 (1998)

\bibitem{colldet} Hersir Sigurgeirsson, Andrew Stuart and Wing-Lok Wan, J. Comp. Phys. {\bf 172}, 766-807 (2001)  

\bibitem{thermsys} D.J. Evans, D.J. Searles and L. Rondoni, e-print cond-mat 0312353

\end{thebibliography}
\end{document}